\def\be{\begin{equation}}
\def\ee{\end{equation}}
\def\bea{\begin{eqnarray}}
\def\eea{\end{eqnarray}}
\def\bse{\begin{subequations}}
\def\ese{\end{subequations}}

\documentclass[prl,twocolumn,showpacs,amsmath,amssymb]{revtex4}
\usepackage{graphicx}
\usepackage{dcolumn}
\usepackage{bm}
\begin{document}
\title{Softer than normal, but not as soft as one might think: Spontaneous flux
       lattices in ferromagnetic spin-triplet superconductors
}
\author{Sumanta Tewari$^{1}$, D. Belitz$^{1,2}$,
        T.R. Kirkpatrick$^{3}$, and John Toner$^{1}$}
\affiliation{$^{1}$Department of Physics and Institute of Theoretical Science,
             University of Oregon, Eugene, OR 97403\\
         $^{2}$Materials Science Institute, University of Oregon, Eugene, OR
         97403\\
         $^{3}$Institute for Physical Science and Technology and Department of
         Physics, University of Maryland, College Park, MD 20742
         }

\date{\today}

\begin{abstract}
A theory is developed for the spontaneous vortex lattice that is expected to
occur in the ferromagnetic superconductors ZrZn$_2$, UGe$_2$, and URhGe, where
the superconductivity is likely of spin-triplet nature. The long-wavelength
fluctuations of this spontaneous flux lattice are predicted to be huge compared
to those of a conventional flux lattice, and to be the same as those for
spin-singlet ferromagnetic superconductors. It is shown that these fluctuations
lead to unambiguous experimental signatures which may provide the easiest way
to observe the spontaneous flux lattice.
%
\end{abstract}


\maketitle

Conventional wisdom holds that ferromagnetism and superconductivity cannot
coexist \cite{Ginzburg_1956, Berk_Schrieffer_1966}. However, recent
experimental studies of UGe$_2$ \cite{Saxena_et_al_2000, Huxley_et_al_2001},
URhGe \cite{Aoki_et_al_2001}, and ZrZn$_2$ \cite{Pfleiderer_et_al_2001}, backed
by band structure calculations \cite{Santi_Dugdale_Jarlborg_2001,
Shick_Picket_2001}, have demonstrated coexistence of both types of order in the
same electron band in these materials. Theoretical considerations suggest that
the superconducting order parameter is of the spin-triplet non-unitary type
\cite{Machida_Ohmi_2001}.

A state which displays coexistence of ferromagnetism and superconductivity is
expected to have many unusual features, among them a spontaneous vortex or flux
lattice. The internal magnetic field generated by the spontaneous magnetization
makes topological excitations, viz. vortices, in the superconducting order
parameter energetically favorable \cite{ground_state_footnote}. Unlike the
well-known Abrikosov flux lattice \cite{Abrikosov_1957}, this {\em spontaneous}
flux lattice state requires no external magnetic field. For spin-singlet
superconductors, such spontaneous flux lattices have been proposed and
theoretically studied previously \cite{Greenside_Blount_Varma_1981,
Ng_Varma_1997, Radzihovsky_et_al_2001}.

\begin{figure}[h]
\includegraphics[width=7.5cm]{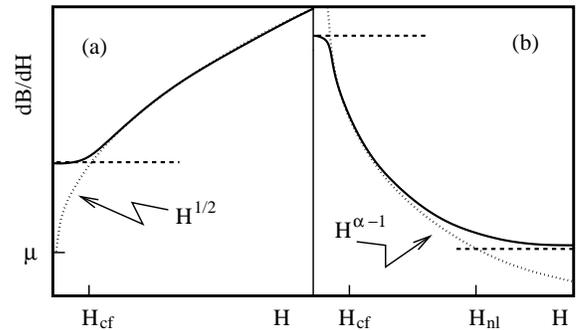}
\caption{\label{fig:1} Schematic representation of the predicted nonanalytic
 behavior of $B(H)$ for clean (a) and disordered (b) systems. $H_{\text{cf}}$ is
 the crystal-field scale that cuts of the nonanalyticity at small fields, and
 $H_{\text{nl}}$ is an upper cutoff scale determined by the disorder. See the text
 for additional information.}
\end{figure}

This Letter addresses the heretofore unstudied problem of spontaneous flux
lattices in spin-triplet p-wave superconductors. One might expect this problem
to be much more complicated than the spin-singlet case, since the order
parameter is much more complicated, which should lead to many more soft modes.
One of our chief conclusions is that, surprisingly, this is not the case.
Rather, the low-energy elastic properties of any hexagonal
\cite{ground_state_footnote_2} spontaneous spin-triplet vortex lattice,
regardless of the precise superconducting order parameter symmetry, map onto
those of the corresponding spin-singlet problem. The reason is that the more
complicated order parameter, while allowing for more modes, also allows for
additional couplings among them, which renders the additional modes massive.
Our second main conclusion is that the very unusual elastic properties of the
spontaneous vortex lattice have easily observable consequences. Specifically,
we predict that in ultraclean samples, the magnetic induction $B$ depends
nonanalytically on an external magnetic field $H$, namely, $B(H) = \mu H +
cH^{3/2}$, with $\mu$ the (linear) magnetic permeability, and $c$ a constant.
The $H^{3/2}$ term is the leading nonanalyticity. In samples with quenched
disorder strong enough to dominate the elastic properties of the flux lattice,
but not strong enough to destroy either the superconductivity or the flux
lattice, the nonanalyticity is the leading term as $H\to 0$, and given by
$B(H)\propto H^{\alpha}$ with $\alpha\approx 0.72$. This is the same result as
in the spin-singlet case \cite{Radzihovsky_et_al_2001}. For very small magnetic
fields either nonanalyticity is cut off by the lattice, which breaks the
spatial rotational invariance, and the disorder-induced nonanalyticity is cut
off at high fields by a field scale, or a corresponding length scale, that
bounds the nonlinear elasticity regime. These results imply the $dB/dH$ versus
$H$ curves shown in Fig.\ref{fig:1}.

To derive these results, we start from a Landau-Ginzburg-Wilson (LGW)
functional that allows for both ferromagnetic and spin-triplet superconducting
order. The superconducting order parameter is a matrix in spin space
\cite{Vollhardt_Woelfle_1990}, $\Delta_{\alpha\beta}(\bm{k}) =
\sum_{\mu=1}^3d_\mu(\bm{k})\left(\sigma_\mu i
 \sigma_2\right)_{\alpha\beta}.$
Here $\alpha$, $\beta$ are spin indices, ${\bm k}$ is the wave vector, and
$\sigma_{1,2,3}$ are the Pauli matrices. This relation expresses the
isomorphism between the complex symmetric matrix $\Delta$ and the complex
3-vector ${\bm d}({\bm k})$. We will consider p-wave symmetry, which implies
$d_{\mu}({\bm k}) = \sum_{j=1}^3 d_{\mu j}\,{\hat k}_j.$ The tensor field
$d_{\mu j}({\bm x})$ is the appropriate order parameter for either He~3 or a
p-wave superconductor. $d_{\mu j}$ is characterized by 18 real numbers, which
explains the rich phenomenology of order parameter textures observed in He~3
\cite{Vollhardt_Woelfle_1990}.

The superconducting part of the LGW functional contains all scalars that can be
constructed from $d_{\mu j}$ and the covariant derivative $D_j = \partial_j - i
q A_j({\bm x})$ ($q=2e/\hbar c$), where the $A_j$ are the components of the
fluctuating vector potential ${\bm A}$, and indices in spin space and real
space, respectively, must be contracted among themselves
\cite{Vollhardt_Woelfle_1990}. Up to bi-quadratic order in covariant gradients
and tensor fields \cite{phi_4_footnote} one finds (with summation over repeated
indices implied)
\bse
\label{eqs:1}
\begin{widetext}
\bea
S_{\text{sc}} &=& \int d{\bm x}\ \Bigl[c_d^{(1)} \left(D_j d_{\mu
l}\right)\left(D_j d_{\mu l}\right)^* + c_d^{(2)}\left(D_j d_{\mu
l}\right)\left(D_l d_{\mu j}\right)^* + c_d^{(3)}\left(D_j d_{\mu
j}\right)\left(D_l d_{\mu l}\right)^* + t_d d_{\mu j}d_{\mu j}^*
\nonumber\\
&&+ u_d^{(1)} d_{\mu j} d_{\mu j} d_{\nu i}^* d_{\nu i}^*
  + u_d^{(2)} d_{\mu j} d_{\mu j}^* d_{\nu i} d_{\nu i}^*
  + u_d^{(3)} d_{\mu j} d_{\nu j} d_{\nu i}^* d_{\mu i}^*
  + u_d^{(4)} d_{\mu j} d_{\nu j}^* d_{\nu i} d_{\mu i}^*
  + u_d^{(5)} d_{\mu j} d_{\nu j}^* d_{\nu i}^* d_{\mu i}
\Bigr].
\label{eq:1a}
\eea
\end{widetext}
The $c_d$, $t_d$, and $u_d$ are the coefficients of the LGW theory.

To describe a ferromagnetic superconductor we need, in addition to $d_{\mu
j}({\bm x})$, a real vector field in spin space, ${\bm M}({\bm x})$, which
serves as the magnetic order parameter. The magnetic part of the LGW functional
takes the form of a $\varphi^4$-theory for the fluctuating magnetization, and
terms for the magnetic field energy and the coupling between ${\bm M}$ and
${\bm A}$ \cite{A_footnote},
\bea
S_{\text{fm}} &=& \int d{\bm x}\ \Bigl[c_m ({\bm\nabla}{\bm M})^2 + t_m{\bm
M}^2 + u_m({\bm M}^2)^2
\nonumber\\
&&+\frac{1}{8\pi}\,\left({\bm\nabla}\times{\bm A}\right)^2
  -{\bm M}\cdot({\bm\nabla}\times{\bm A})\Bigr].
\label{eq:1b}
\eea

Finally, there is a direct coupling between the magnetic and superconducting
order parameters (in addition to the indirect one via the vector potential).
Considerations analogous to those that lead to Eq.\ (\ref{eq:1a}) show that up
to quartic order in the fields there are two such terms,
\be
S_{\text{fm}-\text{sc}} = \int d{\bm x}\ \Bigl[-ig_1
\epsilon_{\mu\nu\lambda}\,M_{\mu}\,d_{\nu i}\,d_{\lambda i}^* + g_2
M_{\mu}\,M_{\nu}\,d_{\nu i}\,d_{\mu i}^*\Bigr],
\label{eq:1c}
\ee
\ese
where $g_1$ and $g_2$ are two additional real coupling constants and
$\epsilon_{\mu\nu\lambda}$ is the Levi-Civita tensor. The complete LGW action
is the sum of the terms in Eqs. (\ref{eqs:1}).

This action is much more complicated than that for an s-wave superconductor.
Nevertheless, the number of soft modes, and their long-wavelength effective
action, are identical to those of the spin-singlet case. In particular, the
spontaneous flux lattice has the same long-wavelength properties. This
surprising result, which is the basis for all of the predictions in this
Letter, is independent of the exact nature of the spin-triplet superconducting
phase. In what follows, we first give a heuristic argument, and then a formal
symmetry argument.

To make the heuristic argument more transparent, let us consider a
superconducting order parameter for the simplest non-unitary state, the
so-called $\beta$-state \cite{Vollhardt_Woelfle_1990}. (Analogous arguments can
be made for other states.) It is given by a tensor product $d =
{\bm\psi}\otimes\,{\bm\phi}$ of a complex vector ${\bm\psi}$ in spin space, and
a real unit vector ${\bm\phi}$ in orbital space, and the ground-state order
parameter is given by ${\bm\psi} = \Delta_0(1,i,0)$, ${\bm\phi} = (0,0,1)$. In
terms of ${\bm\psi}$ and ${\bm\phi}$, the parts $S_{\text{sc}}$ and
$S_{\text{fm-sc}}$ of the action read
\bse
\label{eqs:2}
\bea
&S&_{\!\!\!\!\text{sc}}^{\!\!\!\!\beta}\! =\! \int d{\bm x}\,\Bigl[
t_{\psi}\left\vert{\bm\psi}\right\vert^2\!\! +
u_{\psi}\left\vert{\bm\psi}\right\vert^4\!\! +
v_{\psi}\left\vert{\bm\psi}\times{\bm\psi}^*\right\vert^2\!\!
+c_{\psi}\left\vert{\bm D}{\bm\psi}\right\vert^2
\nonumber\\
&&\hskip -15pt  + \vert{\bm\psi}\vert^2\!\left(
c_{\psi}\left({\bm\nabla}{\bm\phi}\right)^2\! +
d_{\psi}\left({\bm\nabla}\cdot{\bm\phi}\right)^2\right)\! +
d_{\psi}\left\vert\left({\bm\phi}\cdot{\bm D}\right){\bm\psi}
   \right\vert^2\Bigr],
\label{eq:2a}
\eea
\be
S_{\text{fm}-\text{sc}}^{\beta} \!= \!\int d{\bm x}\, \Bigl[-ig_1 {\bm
M}\cdot\left({\bm\psi}\times{\bm\psi}^*\right) + g_2\left\vert{\bm
M}\cdot{\bm\psi}\right\vert^2\Bigr].
\label{eq:2b}
\ee
\ese
The coefficients are simply related to those in Eqs.\ (\ref{eqs:1}).

Compared to a spin-singlet s-wave superconductor, with a complex scalar order
parameter field, we have more order parameter components, and also additional
coupling terms. It turns out that the effects of these two complications
effectively cancel each other, since the additional couplings generate masses
for the extra degrees of freedom, leaving the number of massless degrees of
freedom, and their long-wavelength Hamiltonian, identical to that for the
spontaneous flux lattice in the spin-singlet case.

This can be seen as follows. We consider a region in parameter space where the
ground state of the LGW functional, Eqs.\ (\ref{eqs:2}), (\ref{eq:1b}) is a
vortex lattice \cite{ground_state_footnote}; i.e., a lattice of parallel lines
which each represent a topological ``winding'' singularity of an overall phase
associated with the complex vector ${\bm\psi}$. As in spin-singlet
superconductors, the gauge couplings implicit in the covariant derivatives in
Eq. (\ref{eq:2a}) force ${\bm B} = {\bm\nabla}\times{\bm A}$ to run along these
vortex lines. Physically, this is the Meissner effect. Fluctuations in the
direction of ${\bm B}$ are therefore massive. The ${\bm B} \cdot {\bm M}$
coupling in Eq. (\ref{eq:1b}) then fixes the direction of ${\bm M}$ to also be
parallel to the flux lines, up to massive fluctuations, and the ${\bm
M}\cdot\left({\bm\psi}\times{\bm\psi}^*\right)$ term in Eq. (\ref{eq:2b}) fixes
the direction and phases of ${\bm\psi}$, up to one free (Goldstone) phase whose
vortex singularities are the flux lines. Finally, the direction of ${\bm\phi}$
is fixed by the gauge couplings in Eq. (\ref{eq:2a}), which force ${\bm\phi}$
to also run along the flux lines. Hence, the entire order parameter structure
is determined (up to massive fluctuations) by the positions of the vortex
lines. The only Goldstone modes in the system are thus the two-component
positional fluctuations ${\bm u}$ of the flux lines relative to some perfect
reference lattice of parallel lines with an arbitrary orientation.

A more general conclusion is reached by formal symmetry arguments. The
superconducting part of the action, Eq. (\ref{eq:1a}), is invariant under a
group SO(3)$\times$SO(3)$\times$U(1) of rotations in spin space, rotations in
real space, and gauge transformations \cite{Vollhardt_Woelfle_1990}. The
couplings in Eq. (\ref{eq:1c}) are invariant under SO(3) co-rotations of the
spin part of the tensor $d_{\mu j}$ and the vector ${\bm M}$, which expresses
the fact that both are elements of the same spin space. The magnetic part of
the action, Eq. (\ref{eq:1b}), is invariant only under co-rotations in spin
space and in real space, so the entire action is invariant under a group
SO(3)$\times$U(1). In a ferromagnetic superconducting state this symmetry is
spontaneously broken to SO(2), and the number of Goldstone modes is given by
dim(SO(3)$\times$U(1)/SO(2)) = 3 \cite{remaining_symmetry_footnote}. These are,
two spin-wave-like modes due to the broken spin rotation symmetry, and one
Anderson-Bogoliubov mode due to the broken gauge symmetry. The latter is
rendered massive by means of the Higgs mechanism, and the former can be
eliminated in favor of vortex-lattice degrees of freedom as shown in Ref.\
\onlinecite{Radzihovsky_et_al_2001}. Notice that, although the triplet
superconducting action is invariant under a much larger symmetry group than its
singlet analog, the symmetry properties of the full action, and hence the
number of Goldstone modes, are the same as in the singlet case, in agreement
with the heuristic arguments given above. This symmetry argument is not tied to
a particular superconducting ground state.

We have corroborated the general arguments given above by expanding the action
to Gaussian order about both the $\beta$-state, and an order parameter
appropriate for the A$_1$-phase in He 3, which is another nonunitary state. The
number and nature of the soft modes found is consistent with the general
arguments given above.

The final effective action for any Heisenberg ferromagnetic p-wave
superconductor \cite{remaining_symmetry_footnote} is therefore the same as for
a ferromagnetic s-wave spin-singlet superconductor. It describes an
Anderson-Bogoliubov mode by means of an O(2) nonlinear sigma model for a phase
$\theta$, generalized spin waves by means of an O(3) nonlinear sigma model for
a unit 3-vector ${\hat{\bm\varphi}}$, and a coupling between the two by means
of the gauge field ${\bm A}$. With LGW coefficients $c$ and ${\tilde a}$ one
thus has
\bea
S_{\text{eff}} \hskip -2pt &=&\hskip -3pt \int d{\bm x}\,\Bigl[\frac{c
\Delta_0^2}{2}\,\left({\bm\nabla} \theta({\bm x}) - q{\bm A}({\bm x})\right)^2
\hskip -1pt + \frac{{\tilde a} M_0^2}{2}\,\left({\bm\nabla}{\hat\varphi}({\bm
x})\right)^2
\nonumber\\
&&+\frac{1}{8\pi}\,\left({\bm\nabla}\times{\bm A}({\bm x})\right)^2
-M_0\,{\hat{\bm\varphi}}({\bm x})\cdot\left({\bm\nabla}\times{\bm A}({\bm
x})\right) \Bigr],
\label{eq:3}
\eea
with $\Delta_0$ and $M_0$ the average amplitudes of the superconducting and
magnetic order parameters, respectively.

We now look for saddle-point solutions of this action that take the form of
vortices \cite{ground_state_footnote}, i.e., where the superfluid velocity
${\bm v} = {\bm\nabla}\theta$ obeys the condition
\be
\frac{{\bm\nabla}\times{\bm v}({\bm x})}{2\pi} = \sum_n\int d \tau\ \frac{d{\bm
r}_n}{d\tau}\,
   \delta({\bm x} - {\bm r}_n(\tau))
\equiv {\bm t}({\bm x}).
\label{eq:4}
\ee
Here ${\bm r}_n({\tau})$ is a parameterized line in $\mathbb{R}^3$ representing
the $n$'th vortex line. By minimizing the action with respect to $\theta({\bm
x})$ and ${\bm A}({\bm x})$ we can express the saddle-point action in terms of
the vortex line degrees of freedom ${\bm t}$ coupled to ${\hat{\bm\varphi}}$
\cite{Radzihovsky_et_al_2001, action_footnote},
\bea
S_{\text{eff}}^{(0)} &=& \frac{\pi}{2q^2}\int d{\bm x}\,d{\bm y}\ V({\bm
x}-{\bm y})\,{\bm t}({\bm x})\cdot{\bm t}({\bm y})
\nonumber\\
&& - \frac{2\pi M_0}{q}\int d{\bm x}\,d{\bm y}\ V({\bm x}-{\bm y})\,{\bm
t}({\bm x})\cdot{\hat{\bm\varphi}}({\bm y})
\nonumber\\
&&\hskip -45pt + \frac{a M_0^2}{2}\!\int\! d{\bm x}\,
\left({\bm\nabla}{\hat\varphi}({\bm x})\right)^2\! + 2\pi
M_0^2\lambda^2\!\int\! d{\bm x}\, \left({\bm\nabla}\cdot{\hat\varphi}({\bm
x})\right)^2.
\label{eq:5}
\eea
Here $\lambda = 1/\sqrt{4\pi c q^2 \Delta_0^2}$ is the London penetration
length, $a = {\tilde a} - 4\pi\lambda^2$, and $V({\bm x}) =
(1/4\pi\lambda^2\vert{\bm x}\vert)\exp(-\vert{\bm x}\vert/\lambda)$ is a
screened Coulomb potential.

The action given by Eq.\ (\ref{eq:5}) can be analyzed as explained in Ref.\
\onlinecite{Radzihovsky_et_al_2001}. The equilibrium state is a hexagonal
vortex lattice \cite{ground_state_footnote_2} described by two-dimensional
lattice vectors ${\bm R}_n = (X_n,Y_n)$. Fluctuations of the vortex lattice are
described by a two-dimensional displacement field ${\bm u}({\bm R}_n,z)$ such
that the vortex lines are given by
\be
{\bm r}_n(z) = \left(X_n + u_x({\bm R}_n,z), Y_n + u_y({\bm R}_n,z), z\right),
\label{eq:6}
\ee
where we use $z$ as the parameter of the line. After integrating out the
generalized spin waves, the fluctuation action to second order in the strain
tensor \cite{strain_footnote}
\be
u_{ij}({\bm x}) = \frac{1}{2}\,\left[\partial_i u_j({\bm x}) + \partial_j
u_i({\bm x}) - \partial_{\alpha} u_i({\bm x})\partial_{\alpha} u_j({\bm
x})\right]
\label{eq:7}
\ee
reads
\be
S_{\text{fluc}} = \frac{1}{2}\int d{\bm x}\ \Bigl[\kappa \left(\partial_z^2
{\bm u}\right)^2 + 2\mu\,u_{ij}\, u_{ij} + \lambda\left(u_{ii}\right)^2\Bigr],
\label{eq:8}
\ee
where the elastic constants $\kappa$, $\mu$ (not to be confused with the
magnetic permeability in Fig.\ \ref{fig:1}), and $\lambda$ (not to be confused
with the penetration depth) can be expressed in terms of the coefficients of
the LGW action, Eq.\ (\ref{eq:4}). Unlike the flux lines in a conventional
Abrikosov flux lattice, which are induced by an external magnetic field, the
flux lines in this system are spontaneously generated, and therefore do not
have a preferred direction. As described for the singlet case in Ref.
\onlinecite{Radzihovsky_et_al_2001}, this rotational invariance leads to an
additional softness in the elasticity of the spontaneous flux lattice, due to
the absence of the usual ``tilt energy'' term proportional to $(\partial_z {\bm
u})^2$. This feature leads to anisotropic Gaussian ${\bm u}$-propagators, where
$k_z^4$ scales as ${\bm k}_{\perp}^2$, with ${\bm k} = ({\bm k}_{\perp},k_z)$
the wave vector. Power counting shows that the Gaussian theory is stable for
all dimensions $d>5/2$. This argument neglects rotational symmetry breaking by
crystal fields, which provide a long-wavelength cutoff to the applicability of
our theory, as illustrated in Fig.\ \ref{fig:1}.

The leading corrections to the Gaussian action are terms of the structure
$\partial_{\perp}u(\partial_z u)^2$ and $(\partial_z u)^4$, respectively. They
lead to a least irrelevant operator with scale dimension $-(d-5/2)$, or to a
wave vector dependence of the elastic constant $\mu(k_{\perp}=0,k_z) = \mu
(1+{\text{const.}}\times k_z^{2d-5})$. By arguments analogous to those given in
Ref. \onlinecite{Radzihovsky_et_al_2001} this leads to a nonanalytic
strain-stress relation, and finally to the nonanalytic dependence of $B$ on $H$
discussed above in the context of Fig. 1(a).

Even more unusual elastic properties result from the presence of quenched
disorder. It was shown in Ref.\ \onlinecite{Radzihovsky_et_al_2001} that
ordinary impurities lead to a random-field term in $S_{\text{fluc}}$ that
couples linearly to $\partial_z {\bm u}$. As a result of the strong
random-field effects, the Gaussian theory becomes unstable for all dimensions
$d<7/2$. A renormalization-group analysis showed that the elastic constants, as
well as the variance of the random field, become singular functions of the wave
number, and the corresponding exponents have been calculated to first order in
an expansion in powers of $\epsilon = 7/2-d$. The resulting non-Hookian, or
nonlinear, elastic properties of the vortex lattice extend up to a length scale
$\xi_{\text{nl}} = \kappa^2/w$, where $w$ is the variance of the disorder
distribution. They lead to the predictions shown in Fig.\ \ref{fig:1}(b), which
provide a way to experimentally observe the spontaneous flux lattice.

An important question is whether there is a parameter range where the disorder
is strong enough to lead to observable anomalous elastic effects, but not so
strong as to destroy the p-wave superconductivity, which is very disorder
sensitive. A necessary condition is $\xi\ll\ell\ll a$, with $\xi$ the
superconducting coherence length, $\ell$ the mean-free path, and $a$ the flux
lattice constant. Let us consider ZrZn$_2$ \cite{UGe2_footnote}, where $\xi
\approx 290$\AA\ \cite{Pfleiderer_et_al_2001}. From the normal-state residual
resistivity, $\rho=0.62 \mu \Omega {\rm cm}$, we estimate $\ell\approx 600 -
1000\,{\text\AA}\gg\xi$. The system is thus sufficiently clean to sustain
p-wave superconductivity. On the other hand, from the value of the ordered
moment, $\mu_{\rm s}=0.17 \mu_{\rm B}$ per formula unit
\cite{Pfleiderer_et_al_2001}, which gives rise to a relatively small
spontaneous magnetic induction, B$\approx 0.03$ T, one estimates
$a\approx\sqrt{\phi_0/B}\approx 2,500$\AA. Here, $\phi_0$ is the flux-quantum.
The above condition is thus fulfilled in ZrZn$_2$.

This work was supported by the NSF under grant Nos. DMR-01-32555 and
DMR-01-32726.


\begin{thebibliography}{24}
\bibitem[{gro({\natexlab{a}})}]{ground_state_footnote}
\bibinfo{note}{We consider only cases where the ground state of the system is a
  flux lattice; other states are possible \cite{Greenside_Blount_Varma_1981}.}
\bibitem[{gro({\natexlab{b}})}]{ground_state_footnote_2}
\bibinfo{note}{We have only shown that a hexagonal lattice leads to a local
  energy minimum for our final effective action, Eq.\ (\ref{eq:6}); the global
  minimum could have a different structure. The number and type of Goldstone
  modes would be the same, but the detailed form of the elastic energy would be
  different for a non-hexagonal lattice. See also Ref.\
  \onlinecite{ground_state_footnote}}.
\bibitem[{phi()}]{phi_4_footnote}
\bibinfo{note}{It is easy to verify that keeping terms of higher than quartic
  power in either order parameter in Eqs. (\ref{eqs:2}) does not affect our
  conclusions.}
\bibitem[{A_f()}]{A_footnote}
\bibinfo{note}{If one is interested in ferromagnetism by itself, the
  fluctuating vector potential has no important effects and is usually omitted
  from the action. For the properties of a ferromagnetic superconductor, on the
  other hand, the coupling of the vector potential to the magnetization is
  crucial, as we will see.}
\bibitem[{rem()}]{remaining_symmetry_footnote}
\bibinfo{note}{We consider only magnetic states which are rotationally
  invariant in the plane perpendicular to the direction of the spontaneous
  magnetization.}
\end{thebibliography}

\end{document}